**The intrinsic (trap-free) transistors based on perovskite single crystals with self-passivated surfaces.**

V. Bruevich,[1] L. Kasaei,[1] S. Rangan,[1] H. Hijazi,[1] Z. Zhang,[1] T. Emge,[2] E. Andrei,[1] R. A. Bartynski,[1] L. C. Feldman,[1] and V. Podzorov.[1,*]

[1] Department of Physics and Astronomy, Rutgers University, Piscataway, New Jersey 08854, USA.

[2] Department of Chemistry and Chemical Biology, Rutgers University, Piscataway, New Jersey 08854, USA.

[*] Corresponding author's e-mail address: podzorov@physics.rutgers.edu

Lead-halide perovskites emerged as novel semiconducting materials suitable for a variety of optoelectronic applications. However, fabrication of reliable perovskite field-effect transistors (FETs), the devices necessary for the fundamental and applied research on charge transport properties of this class of materials, has proven challenging. Here we demonstrate high-performance perovskite FETs based on epitaxial, single crystalline thin films of cesium lead bromide ($CsPbBr_3$). An improved *vapor-phase epitaxy* has allowed growing truly large-area, atomically flat films of this perovskite with excellent structural and surface properties. FETs based on these $CsPbBr_3$ films exhibit textbook transistor characteristics, with a very low hysteresis and high intrinsic charge carrier mobility. Availability of such high-performance devices has allowed the study of Hall effect in perovskite FETs for the first time. Our magneto-transport measurements show that the charge carrier mobility of $CsPbBr_3$ FETs increases on cooling, from ~ 30 $cm^2V^{-1}s^{-1}$ at room temperature, to ~ 250 $cm^2V^{-1}s^{-1}$ at 50 K, exhibiting a band transport mostly limited by phonon scattering. The epitaxial growth and FET fabrication methodologies described here can be naturally extended to other perovskites, including the hybrid ones, thus representing a technological leap forward, overcoming the performance bottleneck in research on perovskite FETs.

Lead-halide perovskites have recently emerged as novel semiconducting materials successfully used in solar cells,[1-2] optically pumped lasing and light-emitting diodes (LEDs),[3-6] as well as radiation detectors.[7] However, while the success of perovskites in these applications is stellar, demonstrating high-performance perovskite field-effect transistors (FETs) has proven challenging, despite the surge of recent efforts (see, e.g.,[8]) and the fact that the first FETs based on hybrid perovskites were proposed more than two decades ago in the pioneering work of C. R. Kagan *et al.*[9] Owing to several factors, the commonly used solution-processing approaches to fabrication of perovskite thin films, while being beneficial for large-area solar cells, become less advantageous in a coplanar FET geometry, where the charge carrier flow at the semiconductor-dielectric interface is exceptionally susceptible to the influence of disorder and impurities. In addition, perovskites are prone to ionic drift at relatively low electric fields, leading to *in-operando* modifications of the material's properties (defect formation, self-doping, decomposition), as well as degradation of contacts, or gate electric-field screening.[10-13] Finally, many perovskites readily interact with ambient moisture and oxygen that reduce the (electro)-chemical device stability. As a result, perovskite FETs reported to date exhibit poor device characteristics and/or low carrier mobilities, signaling a performance dominated by disorder, impurities, and device instabilities.[8]

The lack of high-performance perovskite FETs is perhaps the main reason Hall effect measurements have not yet been reported in these devices. Hall effect was demonstrated only in ungated perovskites, where charge carriers are present due to the system's metallicity,[14] an unintentional doping of insulating perovskites,[10] or photogenerated as in photo-Hall effect measurements.[15-17] To fully understand the intrinsic charge transport properties of perovskites requires the development of high-performance FETs, exhibiting both reliable operation and enabling Hall effect investigation, to extract physically meaningful charge carrier mobility.[18]

In developing high-performance perovskite FETs, we built upon the prior pioneering works on the chemical vapor deposition (CVD) of crystalline $CsPbBr_3$ on mica that resulted in epitaxial microplatelets,[5, 19] and nanowires,[20-21] followed up by the growth of more continuous films on mica,[22] sapphire,[23] and strontium titanate.[24] Unlike liquid-phase crystallization[8, 25-26] or high-vacuum evaporation,[27-28] vapor-phase epitaxy in a gas stream can provide both purification and long-range, flat crystalline morphology, essential for charge transport in FETs. Here we have modified and optimized the previously reported vapor-phase epitaxy technique,[22] which has allowed us growing high-quality $CsPbBr_3$ films with atomically flat surfaces over macroscopically large areas, suitable for FET fabrication. The excellent structural quality and chemical purity of our films are confirmed by comprehensive structural and surface analytical studies. FETs fabricated on these films demonstrate nearly perfect *p*-type FET characteristics, albeit with some positive onset voltage due to interfacial doping. Despite the very low voltage sweep rates used in our measurements ($dV/dt = 1$ V·s$^{-1}$), the hysteresis in these FETs is very small. Accordingly, Hall effect measurements yield carrier mobilities comparable to FET mobilities, with both the techniques revealing a temperature dependence of the mobility consistent with a band transport limited by phonon scattering in the range of 50 - 320 K.

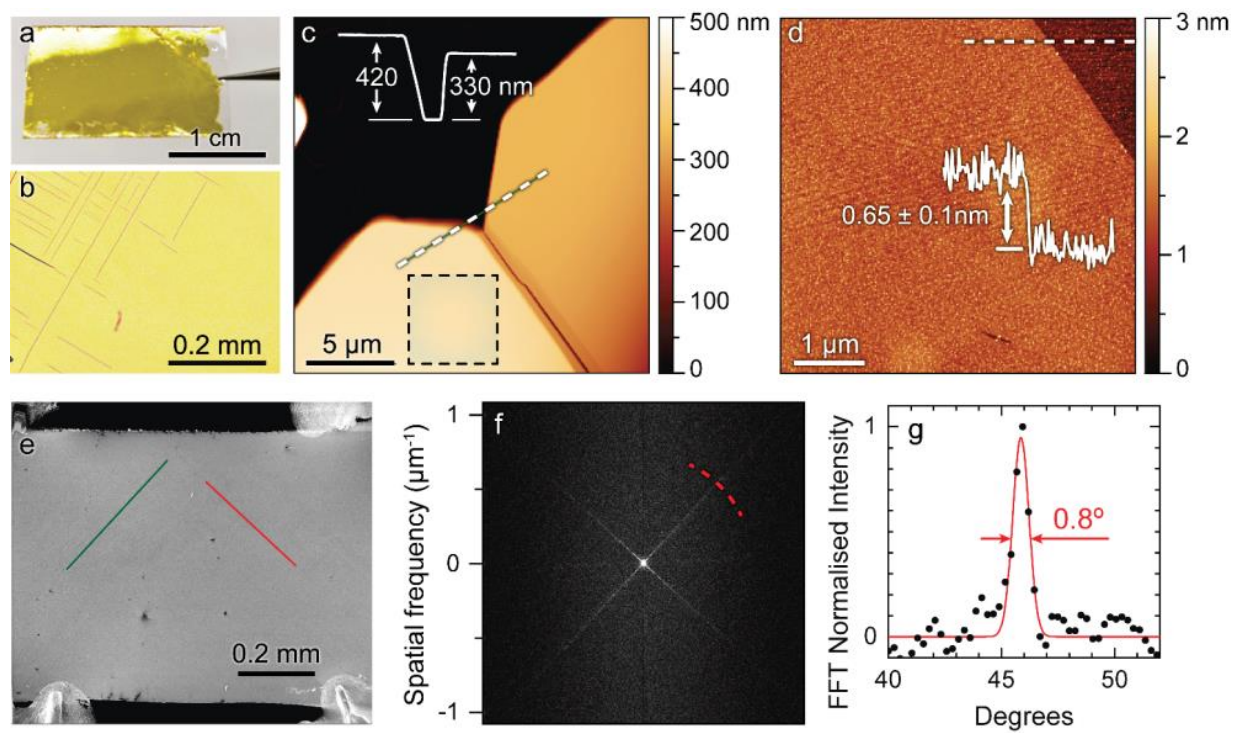

**Figure 1.  Epitaxial single-crystal CsPbBr₃ thin films.  (a)** A photograph of a large-area crystalline CsPbBr$_3$ film grown on mica.  **(b)** An optical microphotograph of an area within a single-crystal grain, showing a perfectly flat region, and a region with fine crystallographically aligned cracks.  **(c)** An AFM topography image collected near the edge of the film (the black region corresponds to the bare mica substrate).  The height profile is taken along the dashed line running through two crystalline grains and the bare substrate.  The dashed box indicates the area of a scan shown on the next panel.  **(d)** An AFM topography scan on a single-crystal grain, showing an atomic step, with a profile taken along the dashed line.  **(e)** A large-scale (low magnification) HeIM secondary electron image of the central part of a CsPbBr$_3$ device with contacts in a Hall-bar/four-probe geometry (the four voltage probes with wires attached can be seen at the corners).  The green and red straight lines indicate the orientation of the faint cross-hatch surface pattern (see text) exhibited by these films.  **(f)** 2D fast Fourier transform of the working area of the CsPbBr$_3$ device shown in the previous panel.  The sharp bright radial lines correspond to the preferred angular orientation of the streaks of the cross-hatch pattern.  **(g)** An angular distribution of 2D-FFT intensity taken along the red dashed line of the previous panel (solid black circles) fitted with a Gaussian (solid red line), with FWHM = 0.8°.

The important design feature of our improved perovskite growth, crucial for the realization of epitaxial $CsPbBr_3$ films with atomically flat surfaces, is the ability to independently fine tune both the position of the tube furnace that defines the temperature profile and the position of the boat containing the material source with respect to the quartz tube holding a stationary substrate (exfoliated mica). The ability to adjust these positions *in-situ* (during the growth) allows dynamic control of the exact temperature distribution in the growth zone and rapid termination of the growth by withdrawing the source (for details, see Methods and Supplementary Information (SI), sec. 1). By using this methodology, we are able to systematically grow large-area, epitaxial crystalline $CsPbBr_3$ thin films of controllable thickness.

We have investigated the morphology, structure and elemental composition of our epitaxial perovskite films by a combination of optical microscopy, atomic-force microscopy (AFM), X-ray diffraction (XRD), X-ray photoelectron spectroscopy (XPS), He-ion microscopy (HeIM), and Rutherford backscattering spectrometry (RBS). The results of these studies confirm the growth of epitaxial, stoichiometric $CsPbBr_3$ films of high crystalline order and chemical purity. These materials analyses are described below, followed by a description of the FET fabrication process, as well as FET and Hall effect measurements.

**Figure 1a** shows a large-scale photograph of the typical epitaxial $CsPbBr_3$ film grown on mica. The morphology of these films correlates with the crystalline grains of mica, showing macroscopically large (typically 1 - 3 mm) single-crystal perovskite domains limited by the grain size of mica. Within single-crystal regions, thicker $CsPbBr_3$ films may develop thin cracks that are always oriented parallel or perpendicular to each other (**Fig. 1b**). **Figure 1c** shows an AFM topography over a $CsPbBr_3$ grain boundary near the edge of the film. The films can be hundreds of nanometers thick and have atomically smooth surfaces. Typically, only a few atomic steps are observed over ~ 10×10 μm$^2$ scan area (**Fig. 1d**). The average height of the atomic steps is 0.65 ± 0.10 nm (inset in **Fig. 1d**), which is consistent with CsPbBr$_3$'s lattice parameter (5.82 Å). The root mean square (rms) roughness of the flat surface of the crystalline areas is smaller than 0.2 nm.

Macroscopically large (a few mm in size), single-crystal domains of our CsPbBr$_3$ films allow fabricating FETs within a single grain, which is essential for avoiding the detrimental effect of grain boundaries on the charge transport and Hall measurements (see, e.g., Ref. [29]) to reveal the *intrinsic* (that is, not dominated by static disorder) transport properties of the material. HeIM was used to explore the CsPbBr$_3$ channel morphology. **Figure 1e** shows a large-scale HeIM image of a millimeter-sized central part of the channel of one of our CsPbBr$_3$ devices with a bare (uncoated) surface. The entire device (including the source and drain contacts) is shown in an optical photograph in SI, **Fig. S2a**. HeIM images of epitaxial CsPbBr$_3$ on mica frequently reveal a very faint cross-hatch texture (as seen more clearly in SI, **Fig. S2b,c**). The exact nature of this contrast has yet to be understood, but it is likely related to small, long-range lateral variations of strain associated with the interaction of epitaxial CsPbBr$_3$ adlayer with the substrate. The long-range character of this texture can be revealed by taking a two-dimensional fast Fourier transform (2D-FFT) of the HeIM images, clearly showing the preferred orientation of the streaks produced by the texture with a very narrow angular distribution (**Figs. 1f,g**). Similar image processing was used in morphological analysis of epitaxial nanowires.[21] The very narrow angular distribution of the streaks of the texture, with the full width at half maximum (FWHM) of 0.8°, shows long-range (device-scale) ordering exhibited by this surface pattern, thus implying a single crystalline nature of the entire macroscopic channel of the device.

The stoichiometry of lead-halide perovskites can vary significantly depending on the precursor materials and growth conditions.[30] To investigate the stoichiometry of our films we have performed XPS characterization. **Figure 2a** shows core levels XPS spectra of an epitaxial single crystalline $CsPbBr_3$ film and that of a cleaved bulk stoichiometric crystallized melt, both prepared from the same stoichiometric mixture of precursors (Methods). XPS spectra of these two types of samples show the same relative elemental composition, with the core-level energies indicating that the chemical environments of each element in the epitaxial film and the crystallized melt are the same (see also SI, sec. 3) and similar to those reported for $CsPbBr_3$.[30-31] As expected, the epitaxial film shows narrower XPS peaks (**Fig. 2a**), which is a signature of a more uniform chemical environment compared to that of the bulky crystallized melt. Because in this work we handle all our samples at ambient conditions (in a regular laboratory air), the surface of the $CsPbBr_3$ films shows some contamination with carbon and oxygen (SI, sec. 3). XPS depth profiling confirmed that small amounts of C and O are only detectable in the top atomic layer of these epitaxial films (**Fig. S4**), and, as a consequence, the XPS core levels spectra of the epitaxial films reflect a bulk environment.

Although the chemical environment can be assessed using XPS, an accurate evaluation of the stoichiometry cannot be achieved due to the inaccuracy of the photoemission cross-sections available for these elements (SI, sec. 3). Therefore RBS measurements have been performed.[32] **Figure 2b** shows a fitted RBS spectrum of an epitaxial $CsPbBr_3$ film on mica. The determined stoichiometry of the film is very close to the nominal one.

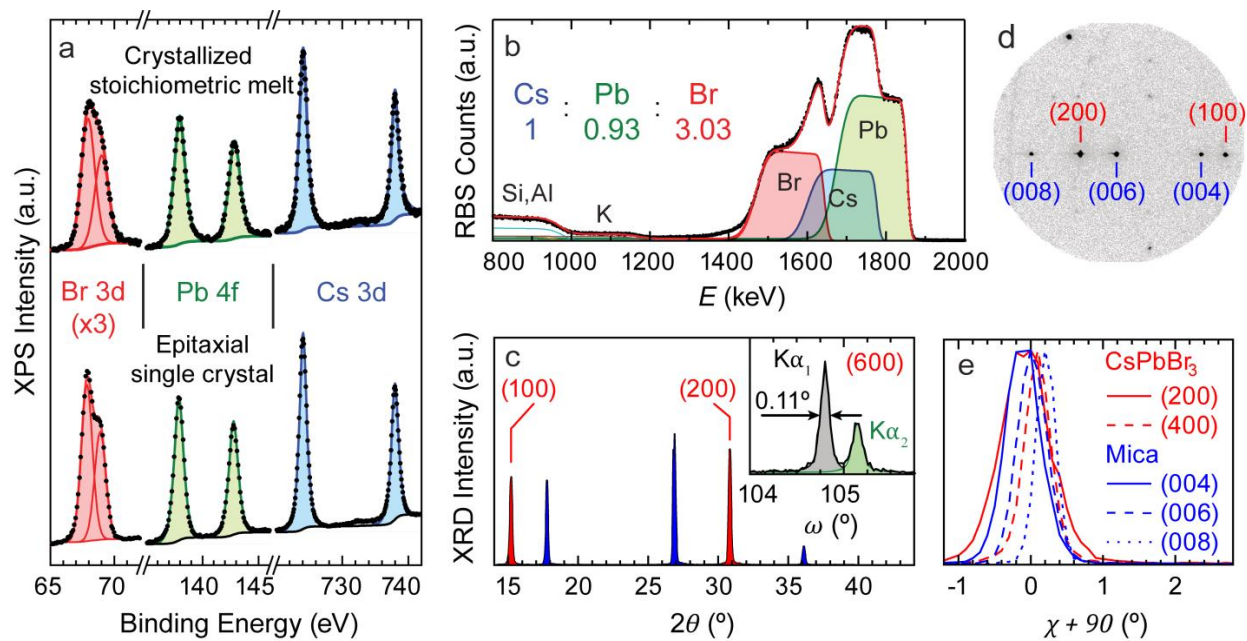

**Figure 2. Compositional and structural characterization of epitaxial single-crystal CsPbBr₃ films.**
(**a**) XPS spectra of a cleaved bulky CsPbBr₃ sample crystallized from a stoichiometric melt (top) and an epitaxial CsPbBr₃ film grown on mica (bottom). (**b**) An RBS spectrum of an epitaxial CsPbBr₃ film on mica (fine black dots), with a theoretical fit (solid red line), showing deconvolution into the individual elements (thin solid curves with colored shading). (**c**) An XRD $\theta$ - $2\theta$ scan of an epitaxial crystalline CsPbBr₃ film on mica (red and blue shaded peaks/indices correspond to CsPbBr₃ and mica, respectively). The inset shows an XRD rocking curve for the (600) reflection, with FWHM = 0.11° for the K$\alpha_1$ peak. (**d**) Area detector image of Bragg reflections of the single-crystal CsPbBr₃ film and the mica substrate. The near-perfect alignment (i.e., all in horizontal line) of the ***a***-axis vector of CsPbBr₃ and ***c***-axis vector of mica confirms an epitaxial growth. (**e**) The $\chi$-scans for different XRD reflections of the epitaxial CsPbBr₃ film and the mica substrate, indicating a high degree of alignment.

To investigate the crystallographic structure and quality of our epitaxial CsPbBr₃ films, we have carried out room-temperature XRD measurements (Methods). **Figure 2c** shows $\theta$ - $2\theta$ XRD scan of the film. The peaks at $2\theta = 15.19°$ and $30.80°$ are the reflections from the (100) and (200) planes of CsPbBr₃, while those at $2\theta = 17.71°$, $26.83°$ and $36.01°$ are the reflections from the (004), (006) and (008) planes of the single-crystal mica. The lattice of our $CsPbBr_3$ film was found to be cubic with a lattice parameter 5.82 Å, consistent with the earlier findings from nanoplatelet growth.[5] Excellent crystalline quality of these films is confirmed by the rocking curve measurements ($\omega$-scan inset in **Fig. 2c**). The FWHM of the (600) plane's reflection at 105.19° ($K\alpha_1$ band) is 0.11°, indicating a very low mosaic spread of the crystal. **Figure 2d** shows the area detector image of Bragg reflections of the $CsPbBr_3$ film on mica. The perovskite's XRD pattern consists of sharp peaks (denoted with red indices) aligned very well with the peaks of mica (denoted with blue indices). Indeed, the measured $\chi$-scans of various perovskite and mica reflections (**Fig. 2e**) show good peak clustering, with all the maxima occurring within $\delta\chi = \pm 0.2°$, confirming a very good in-plane alignment between the mica and the epitaxial $CsPbBr_3$ adlayer. Such a nice alignment of the peaks of the adlayer and the substrate is a very good indicator of epitaxial growth.

**Figure 3a** shows a sketch of our $CsPbBr_3$ FET structure, employing carbon contacts, a *parylene* gate dielectric, and a thermally evaporated silver gate (Methods). To limit the undesired poling effect (SI, sec. 4), we carried out measurements of the output characteristics (that is, the source-drain current as a function of the source-drain voltage, $I_{SD}(V_{SD})$, recorded at a fixed gate voltage, $V_G$) of our perovskite FETs at a reduced temperature (200 K) and in a relatively small range of $V_{SD}$ (**Fig. 3b**). Despite the very low voltage sweep rate ($dV_{SD}/dt = 1$ V·s$^{-1}$), almost no hysteresis is observed (inset in **Fig. 3b**).

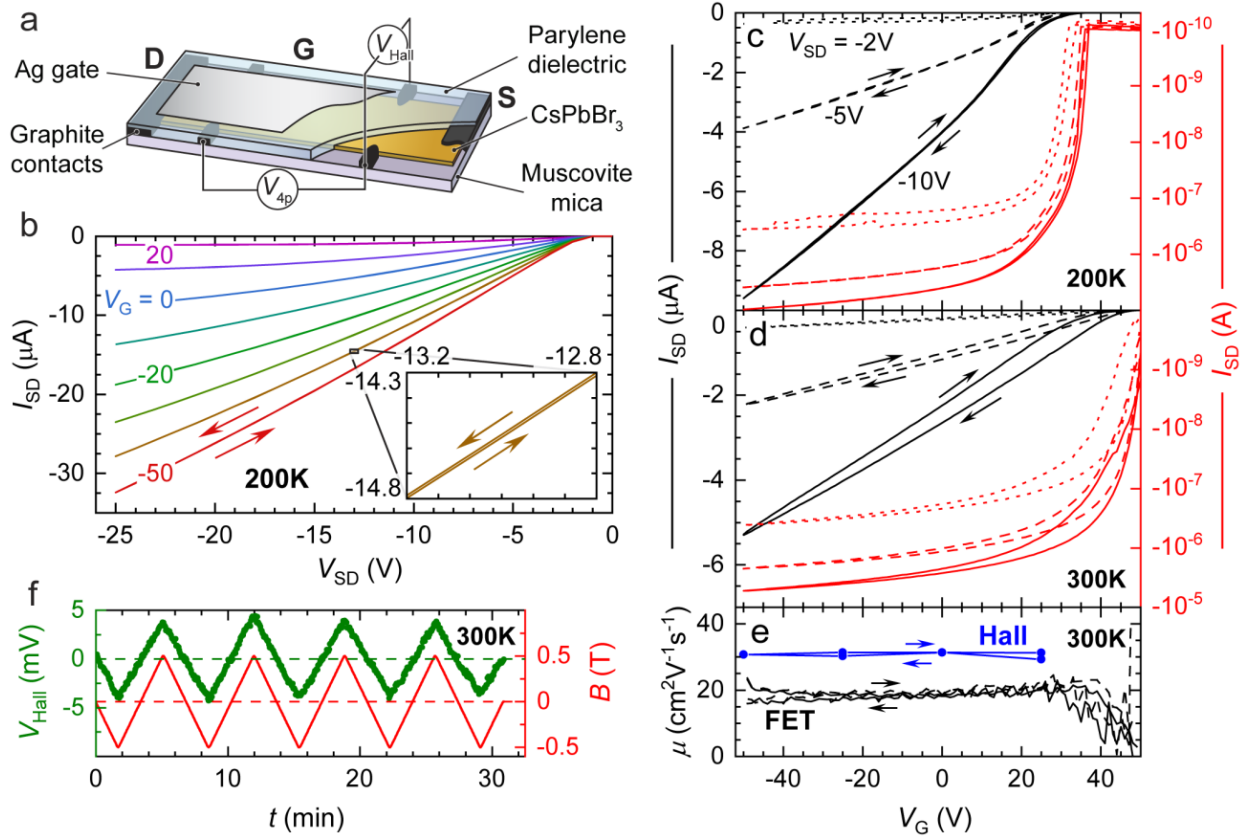

**Figure 3. Electrical characterization of high-performance epitaxial single-crystal CsPbBr₃ FETs.**
(a) A sketch of the FET structure. (b) Output characteristics of a CsPbBr$_3$ FET recorded at 200 K and $V_G$ in the range between 20 and - 50 V (with a step of 10 V). The inset is a zoom-in showing a negligible hysteresis even at a very low sweep rate ($dV_{SD}/dt = 1$ V·s$^{-1}$) used here. (c, d) Transfer characteristics of a CsPbBr$_3$ FET recorded at different $V_{SD}$ (as indicated) at 200 and 300 K. The gate voltage sweep rate was $dV_G/dt = 1$ V·s$^{-1}$. (e) The FET's linear-regime four-probe mobility, $\mu_{FET}$ (thin black solid and dashed lines, corresponding to $V_{SD} = - 5$ and - 10 V, respectively), and the Hall mobility, $\mu_{Hall}$ (blue solid symbols), as a function of $V_G$. (f) An example of Hall voltage measurements in a CsPbBr$_3$ FET, showing $V_{Hall}$ (green dots) registered as the $B$ field (solid red line) is ramped up and down between -0.5 and 0.5 T. All the FET measurements are performed by slowly sweeping $V_{SD}$ or $V_G$ back and forth (at 1 V·s$^{-1}$) to show hysteresis (the arrows indicate the direction of the sweep). The reliability factor for the linear-regime FET mobility extraction from panels **c** and **d**, calculated according to Ref. [18], is $r_{lin} = 100 \pm 1$ %. Panels **c**, **d**, **e** and **f** represent the same device with a parylene-*F* gate dielectric.

The channel conductivity and mobility were assessed via the gated four-probe measurements from the linear-regime FET transfer characteristics (that is, $I_{SD}(V_G)$ dependences recorded at fixed $V_{SD}$ ($|V_{SD}| < |V_G|$), **Fig. 3 c, d**) and the corresponding four-probe voltage, $V_{4p}(V_G)$ (see Methods and Ref. [33]). The four-probe measurements are necessary for addressing a potential problem of contact resistance and thus correctly evaluating the mobility. The ability to fabricate long-channel perovskite FETs (with the channel lengths $L \geq 2$ mm) is helpful in minimizing the detrimental poling effect, since, with long $L$, a device operation not dominated by contact resistance can be maintained even at a relatively low $|V_{SD}| \leq 10$ V. Indeed, as previously shown for organic FETs, a relatively high $V_{SD}$ is normally required to suppress the Schottky contact resistance.[34] A positive onset voltage in the transfer characteristics of our perovskite FETs (**Fig. 3 c, d**) originates from a channel doping by the *parylene* gate dielectric grown directly at the surface of CsPbBr$_3$ (SI, sec. 5). The gate leakage current in all our FETs was negligible (not more than 0.003% of $I_{SD}$). **Figures 3 c, d** show that the channel conductivity of CsPbBr$_3$ FETs is effectively modulated with $V_G$. By applying a sufficiently high positive $V_G$, the transistor can be turned off completely (its channel depleted), leading to the on-off ratio of about $10^5$. The linear-regime FET mobility, $\mu_{FET}$, measured via the four-probe technique, is $V_{SD}$- and $V_G$-independent and exhibits negligible hysteresis (**Fig. 3e**), showing that $\mu_{FET}$ is independent of the carrier density. Except for exhibiting a noticeable onset voltage, the epitaxial CsPbBr$_3$ FETs with parylene gate dielectric essentially operate as classic, textbook *p*-type FETs.

Despite the very low sweep rates used in all our measurements ($dV_G/dt = dV_{SD}/dt = 1$ V·s$^{-1}$), hysteresis in our FETs is small even at room temperature and virtually indiscernible below 250 K (**Fig. 3**).  In contrast, characteristics of most perovskite FETs reported in the literature to date are far from ideal (see, e.g., Ref. [8]).  In many cases, mobilities are extracted from highly non-linear device characteristics, ignoring such issues as contact resistance, overwhelming hysteresis, and rapid device degradation, leading to limited result reliability.  Although approaches based on blending multiple cations or washing the films with orthogonal solvents were able to partially remedy some of these issues, the resultant mobilities in solution-processed polycrystalline perovskite FETs remain limited.[35]

We believe the striking improvement in device performance presented here is mainly due to the three factors.  First, fully inorganic $CsPbBr_3$ was reported to be more thermally stable than its hybrid cousins.[36]  Second, the use of high-quality single crystalline perovskite films is crucial for FET performance.  And third, the maximum longitudinal electric field applied in our measurements ($V_{SD}/L \approx 50$ V·cm$^{-1}$) is far lower than those typically reached in earlier reports (> 5000 V·cm$^{-1}$),[35] which helps minimizing undesired ionic drift.

Having fabricated high-quality perovskite FETs, we were able to measure the Hall effect in these devices.  **Figure 3f** shows an example of a Hall voltage, $V_{Hall}$, measured in one of our epitaxial $CsPbBr_3$ FETs (green dots) at 300 K, as the magnetic field $B$ is slowly ramped between -0.5 and 0.5 T at a rate d$B$/d$t$ = 0.005 T·s$^{-1}$ (solid red line), revealing a $V_{Hall}$ signal with a high signal/noise ratio ($\approx 200$), thus allowing a high-precision measurement of the Hall mobility, $\mu_{Hall}$, of the FET (see also SI, sec. 6).  To the best of our knowledge, this is the first Hall effect measurement in a perovskite FET, including hybrid or all-inorganic perovskites, 2D or 3D systems, lead-based or lead-free compounds.

**Figure 3e** compares the Hall mobility and the four-probe linear-regime FET mobility, $\mu_{Hall}$ and $\mu_{FET}$, measured in the same $CsPbBr_3$ FET.  Although both $\mu_{Hall}$ and $\mu_{FET}$ are almost independent of $V_G$ and thus of the carrier density, in this particular device, these mobilities differ by a factor of ~ 1.5. However, we emphasize that this difference is not of an intrinsic nature, as it slightly varies from device to device. For example, the device used in the $\mu(T)$ measurements (**Fig. 4**) is closer to ideal, with $\mu_{Hall} \approx \mu_{FET}$ at room temperature. In principle, differences between $\mu_{Hall}$ and $\mu_{FET}$ in FETs can be associated with the presence of morphological defects (cracks or grain boundaries) in the channel,[29] a coexistence of hoping and band carriers (SI, sec. 7),[37] or the effect of carrier scattering by charged impurities.[38] These mechanisms however are not understood in perovskites at all, and the high-performance FETs developed here could become an effective tool for their systematic investigation.

To better understand the fundamental mechanisms governing the charge carrier mobility in epitaxial CsPbBr$_3$ FETs, we have carried out temperature variable measurements of the FET and Hall mobilities, $\mu_{FET}(T)$ and $\mu_{Hall}(T)$, of these devices (**Fig. 4**). The mobility of devices with crack-free channels is typically $\mu_{FET} \approx \mu_{Hall} \approx 30$ cm$^2$V$^{-1}$s$^{-1}$ at ~ 300K and increases with decreasing temperature. On cooling below 100 K, CsPbBr$_3$ films gradually develop microscopic cracks, as identified by *in-situ* monitoring of $I_{SD}$, or microscopy performed after warming (SI, sec. 8). Such micro or nano cracks seem to affect $\mu_{Hall}$ to a greater degree than $\mu_{FET}$, which is reminiscent of the effect of grain boundaries on the Hall mobility recently investigated by Choi *et al*. in polycrystalline organic transistors.[29] Similar effects might contribute to the difference between $\mu_{Hall}$ and $\mu_{FET}$ observed here at low temperature (**Fig. 4**).

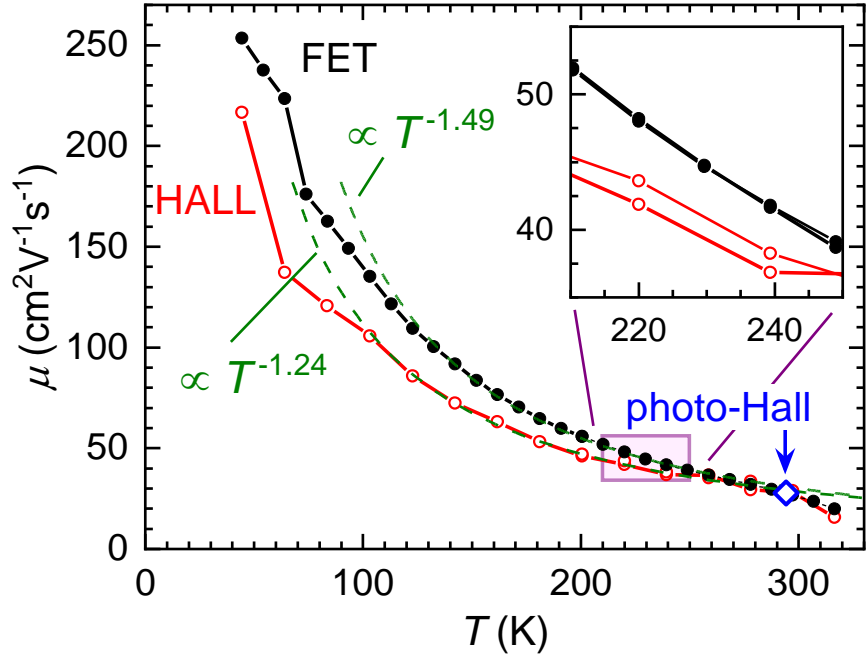

**Figure 4. The temperature dependence of the mobility of epitaxial single-crystal CsPbBr₃ FETs.** The four-probe $\mu_{FET}$ and $\mu_{Hall}$ (solid black and open red symbols, respectively) were measured in a CsPbBr₃ FET operating in the linear regime. The room-temperature photo-Hall mobility measured in the same CsPbBr₃ film (before the gate dielectric and the gate electrode were deposited) is shown for comparison (an open blue diamond) (SI, sec.9). The dashed lines are the fits with $\mu(T) = \text{const} \cdot T^{-\gamma}$, with the exponent $\gamma$ indicated. The reversibility of these measurements was verified by temperature cycling and making sure that $\mu(T)$ reproduces (Methods). The inset is a zoom-in showing a very minor hysteresis in the $\mu(T)$ dependences.

Both $\mu_{Hall}$ and $\mu_{FET}$ monotonically increase on cooling down to ∼ 50 K, with $\mu_{Hall}(T)$ and $\mu_{FET}(T)$ dependences well fitted with a $T^{-\gamma}$ function with the exponent $\gamma$ between 1.24 and 1.49, as shown in **Fig. 4** with the dashed lines. This behavior suggests that acoustic-phonon scattering is the dominant mechanism limiting the charge carrier mobility of the system.[39-42] Such an assignment, however, may be considered an oversimplification, because of the model requirement of a purely parabolic single-carrier band necessary to arrive to the $T^{-1.5}$ dependence.[43] Although such purely parabolic bands and the absence of other (competing) scattering mechanisms are rarely encountered in practice, density functional theory (DFT) calculations predict that the non-parabolicity factor in $CsPbBr_3$ is close to unity.[44] At low temperatures ($T < 100$ K), the experimental $\mu(T)$ dependence is affected by formation of micro/nano cracks (SI, sec. 8), thus showing some notable deviations from the $T^{-\gamma}$ behavior. Nevertheless, both Hall and FET mobilities are monotonically growing with cooling from 320 K down to 50 K, signifying a phonon dominated (rather than defect and impurity dominated) charge carrier scattering regime. Overall, the FET and Hall effect behaviors observed here are consistent with the high-quality, band-semiconductor FETs exhibiting nearly 100% *intrinsic* (static disorder free) band transport.

The charge transport in FETs occurs in proximity to the semiconductor-dielectric interface, and it can thus be affected by the interface's morphology, purity, as well as interaction of the carriers with the gate dielectric.[45] To compare the interfacial carrier mobility, $\mu_{FET}$, extracted from $CsPbBr_3$ FETs with the bulk mobility of our films, we have carried out photo-Hall effect measurements in ungated devices based on the same epitaxial films (for details on *ac* photo-Hall technique, see SI, sec. 9 and Refs. [15, 46]). The room-temperature $\mu_{photo\text{-}Hall}$ value obtained in these measurements (an open blue diamond in **Fig. 4**) is similar to $\mu_{FET}$ and $\mu_{Hall}$ measured in the FETs. Such a match and the observed strong $\mu(T)$ dependence are good signatures of the interfacial charge transport in our epitaxial $CsPbBr_3$ FETs not being surface/interface limited, but reflecting the true intrinsic transport properties of the nearly ideal crystalline semiconductor.

In covalent 3D inorganic semiconductors, deep traps associated with dangling bonds at surfaces have to be passivated before working FETs can be fabricated.[47] The development of surface passivation techniques for silicon has led to the breakthroughs, marking the beginning of modern semiconductor electronics.[48]  Unlike covalent 3D semiconductors, van der Waals molecular crystals intrinsically lack surface dangling bonds, which has been critical for their successful integration in organic FETs.[49-53]  Analogously, layered inorganic (semi)conductors, the precursors of monolayer materials, including transition-metal dichalcogenides and graphene, also lack dangling bonds at their exfoliated surfaces, which has been instrumental in the realization of the first working FETs based on these materials achieved in regular ("dirty") laboratory settings simply by using Scotch tape exfoliation and without special surface passivation techniques.[54-55]  Given the rich history of semiconductor scientists' struggle with surface defects, it is remarkable that high-performance FETs can be created on as-grown surface of a 3D CsPbBr$_3$ material, without any additional surface passivation and with all the sample handling done in regular laboratory air.  Indeed, based on the results presented here, we can confidently conclude that the surface properties of our as-grown CsPbBr$_3$ films are not dominated by traps.  Such a remarkable behavior is due, in part, to the fact that the epitaxial growth utilized here leads to an atomically smooth surface of CsPbBr$_3$ films, making it perfectly suitable for top-gate FET fabrication.

It is also worth noting that the epitaxial CsPbBr$_3$ FETs reported here exhibit no degradation over a long-term storage in air (as tested for up to several months) or during extended electric measurements (that sometimes run for several weeks on end), with the exceptions of driving the FET into a saturation regime or micro crack formation at low temperatures.

In summary, we have demonstrated truly high-performance perovskite FETs based on CsPbBr$_3$, exhibiting textbook FET characteristics and excellent stability.  The outstanding quality of these devices is largely due to the atomically flat, single crystalline morphology of epitaxial CsPbBr$_3$ films achieved over macroscopic length scales, as confirmed by comprehensive structural, morphological and surface analysis. The detailed Hall effect and FET measurements reveal the intrinsic band transport occurring in these devices, with the mobility limited by phonon scattering. This work suggests that epitaxial perovskites emerge as a perfect platform for the fundamental studies of charge transport phenomena in this important class of materials. The robust, simple, and effective method for fabricating perovskite FETs that exhibit stable and ultimately efficient intrinsic charge transport paves a way for a plethora of new perovskite-based devices, such as light emitting FETs, electrically pumped injection lasers, better radiation detectors, sensors and memories, to name a few.

## Methods.

**1. Vapor-phase epitaxy of CsPbBr$_3$ films and crystallization of bulky samples from a melt.** Cesium bromide (CsBr) and lead(II) bromide (PbBr$_2$) precursors of 99.999% purity (Sigma-Aldrich) were mixed at 1:1 molar ratio and reacted by heating this stoichiometric mixture in an alumina boat at 380 $^{o}$C for 12 h in inert atmosphere. Mineral muscovite mica (Electron Microscopy Sciences) was cleaved in a regular laboratory air just before the growth. The alumina boat with the source material and the mica substrate were loaded to a custom-designed CVD growth reactor consisting of a quartz tube in a tube furnace (see SI, sec. 1 for details), with UHP He used as a carrier gas. Before the growth, the apparatus was evacuated and purged with the gas for several times, after which the gas flow and pressure were stabilized at 100 sccm and 0.1 bar, respectively. The source was then melted by setting its temperature to 560 $^{o}$C. The substrate placed downstream is initially heated to just above 500 $^{o}$C, which results in sublimation of possible contaminants and undesired deposits off its surface, thus cleaning it. After stabilizing the temperature, the epitaxial growth of crystalline CsPbBr$_3$ on mica is initiated by shifting the furnace and the source with respect to the quartz tube by 1 inch in the direction opposite to the gas flow, which slightly reduces the substrate's temperature. Typical duration of the growth was 15 min. The growth was then abruptly terminated by withdrawing the source from the hot zone of the furnace without changing the position of the substrate with respect to the temperature distribution. Subsequently, the furnace is turned off, and the substrate gradually cools down to room temperature. The bulky stoichiometric CsPbBr$_3$ crystals were prepared by quickly melting and cooling the exact stoichiometric mixture of the precursors (1:1 molar ratio), which resulted in a polycrystalline solid CsPbBr$_3$. The correct stoichiometry of the melt-crystallized samples is ensured by weighing the precursors with accuracy better than 0.1 %.

**2. AFM** measurements were performed using NT-MDT SolverNEXT microscope in a tapping mode.

**3. XPS** studies were performed using charge compensation, on either a K-Alpha or an ESCALAB 250xi Thermo instrument, with an overall energy resolution of 0.7 eV. Adventitious C 1s is measured at 284.8 eV. A mechanism allowed for in-situ sample cleavage/surface removal in the ESCALAB 250xi.

**4. RBS** studies were performed using a beam of $He^{++}$ ions from the standard 1.7 MV IONEX tandem accelerator. The He beam energy and diameter were 2 MeV and 2 mm, respectively. The typical ion-beam current on the sample during measurements was 10 nA. The RBS spectra were recorded using a computer-based data collection system and analyzed using the SIMNRA simulation program (downloadable at https://www2.ipp.mpg.de/~mam/) with the corresponding Rutherford cross-sections.

**5. HeIM** secondary electron images of epitaxial $CsPbBr_3$ films were recorded using the Zeiss Orion Plus helium ion microscope at an acceleration voltage of 30 kV. During imaging, the chamber pressure was at $3x10^{-7}$ Torr, and the typical ion beam current was 1.3 pA. While capturing large-scale images of macroscopic devices (**Fig. 1e**), the field-of-view was 950 µm.

6. **XRD** measurements were performed with a Bruker Vantec-500 area detector and a Bruker FR571 rotating-anode x-ray generator operating at 40 kV and 45 mA and equipped with a 3-circle Azlan goniometer. The system used 0.5 mm pinhole collimation and a Rigaku Osmic parallel-mode (e.g., primary beam dispersion less than $0.04^o$ in $2\theta$) mirror monochromator (Cu Kα; $\lambda$ = 1.5418Å). Data were collected at room temperature (20 C). Spatial calibration and flood-field correction for the area detector were performed at detector distance used prior to data collection. The 2048x2048 pixel images were collected at a fixed detector ($2\theta$) angle. Data collection and rocking curve creation: Bruker GADDS v.4.1.51 (2015). Data display and graphics: MDI JADE7 v.7.0.6 (2004).

**7. Fabrication of CsPbBr$_3$ FETs.** All sample handling (patterning CsPbBr$_3$ films, contact deposition, etc.) was done in a regular laboratory air under an optical zoom microscope (Leica). The epitaxial CsPbBr$_3$ films on mica were patterned using a sharp razor to select homogeneous, crack-free, single crystalline regions of several mm in size. Contacts were hand painted with isopropanol-based graphite paint (EMS), and 25 µm-thick silver wires were attached to the contact pads. A parylene conformal coating was used to grow a polymer gate dielectric of a thickness in the range 1 - 2 µm (for details, see Refs. [34, 56]), which is much greater than the roughness of parylene's top surface (~ few nm), thus ensuring a uniform gate electric field in the accumulation channel of top-gate FETs. The thickness of parylene was measured with a UV-vis reflectance spectrometer (SemiconSoft). The two types of parylene (*N* and *F*) were used in our CsPbBr$_3$ FETs, and both resulted in high-quality working transistors. The structure was then topped with a 30 nm-thick silver gate thermally evaporated through a shadow mask. Precise geometry of devices, including the channel width (*W*), length (*L*), and the distance (*D*) between the voltage probes of the four-probe contact structure were determined from optical microphotographs.

**8. Electrical characterization of CsPbBr$_3$ FETs.** Measurements of CsPbBr$_3$ FETs were performed after mildly poling the devices at $V_{SD}$ = - 10 V and $V_G$ = 0 for 24 h in vacuum (at a pressure < 1 mTorr). FETs' characteristics were then recorded using a pair of Keithley-2400 source-meters, with one of them connected in the source-drain loop, and the other one connected between the (grounded) drain and the gate. For the four-probe measurements, a Keithley-6514 electrometer was additionally connected to the pair of longitudinal voltage probes (**Fig. 3a**). The FET carrier mobility, $\mu_{FET}$, was calculated from the dependence of the four-probe channel conductivity on the gate voltage, $\sigma_{4p}(V_G)$, recorded in the linear regime of FET operation. The four-probe FET mobility as a function of the gate voltage was calculated using the formula:

$$\mu_{FET}(V_G) = \frac{1}{C_i} \cdot \frac{D}{W} \cdot \frac{d\sigma_{4p}}{dV_G},$$

where $C_i$ is the gate-channel capacitance per unit area. For more details on the four-probe mobility measurements in FETs and potential pit-falls, see Ref. [33].

**9. $dc$-Hall effect measurements in CsPbBr$_3$ FETs** were carried out in a magnetic field of up to $B = \pm 0.55$ T generated by an electro-magnet (GMW, 5403 Dipole, 76 mm). The magnetic field was calibrated with a Gauss/Tesla meter (F. W. Bell, model 5180). The Hall voltage was measured with a Keithley-6514 electrometer. To correctly measure the true Hall voltage, $V_{Hall}$, measurements were performed in both polarities of the $B$ field (by sweeping it up and down) to ensure the correct sign reversal of the Hall voltage. In addition, a slowly drifting parasitic background was subtracted via a polynomial fit (SI, sec. 6). The Hall mobility was then calculated as:

$$\mu_{Hall} = \frac{D}{W} \cdot \frac{1}{V_{4p}} \cdot \frac{\partial V_{Hall}}{\partial B}.$$

During these measurements, FETs were kept in the linear regime at all times. For more details on $dc$-Hall effect measurements in FETs, see, e.g., Ref. [51] and Refs. therein.

**10. Temperature variable measurements** were carried out in a closed-cycle He cryostat (Advanced Research Systems). $\mu(T)$ measurements were performed at stabilized temperature set points in the range 50 - 320 K. Reversibility of $\mu(T)$ dependence has been verified by cycling the temperature between 200 and 320 K and repeating the measurements. Below 100 K, the CsPbBr$_3$ films were developing micro cracks leading to gradual device degradation (SI, sec. 8).

**11.** *ac* **photo-Hall effect measurements in ungated epitaxial crystalline CsPbBr$_3$ films** were carried out using a highly sensitive *ac*-Hall technique.[46]  The sample (unencapsulated CsPbBr$_3$ film with contacts) was placed in a hermetic chamber back-filled with a UHP Ar gas.  A calibrated 458 nm LED (OSRAM, LZ4-V4MDPB-0000) was used as light source.  The frequency of the ac-*B* field was 0.2 Hz.  The measured photo-Hall mobility was found to be nearly light intensity independent, similarly to the earlier photo-Hall effect measurements in solution-growth bulk lead-halide perovskite crystals,[15-16] as well as single crystalline organic semiconductor rubrene.[46]

**Acknowledgements.**

The authors are grateful to the following programs for the financial support of this work. V.B. and V.P. acknowledge support from the National Science Foundation under the grant ECCS-1806363 and the Rutgers Initiative for Materials Research (iMR) seed grant 300761.

**Author contributions.**

V.B. performed epitaxial perovskite growth, FET fabrication, and transport measurements. L.K., H.H. and L.C.F. performed HeIM imaging and RBS measurements with the corresponding data analysis. S.R. and R.A.B. performed XPS measurements and analysis.  T.E. performed XRD characterization.  V.B., Z.Z. and E.A. performed AFM characterization.  V.B. and V.P. designed the experiments, analyzed the charge transport data, and wrote the manuscript with input from all authors.

**Supplementary Information**

9/10/2021

**The intrinsic (trap-free) transistors based on**

**perovskite single crystals with self-passivated surfaces.**

V. Bruevich,[1] L. Kasaei,[1] S. Rangan,[1] H. Hijazi,[1] Z. Zhang,[1] T. Emge,[2] E. Andrei,[1] R. A. Bartynski,[1]

L. C. Feldman,[1] and V. Podzorov.[1,*]

[1] Department of Physics and Astronomy, Rutgers University, Piscataway, New Jersey 08854, USA.

[2] Department of Chemistry and Chemical Biology, Rutgers University, Piscataway, New Jersey 08854, USA.

[*] Corresponding author's e-mail address:  podzorov@physics.rutgers.edu

### 1. Vapor-phase epitaxial growth of single crystalline CsPbBr₃ films.

**Figure S1** shows a sketch of the epitaxial vapor-phase growth apparatus used in this work. The growth reactor consists of a quartz tube (ID = 20 mm, OD = 25 mm) placed in the Lindberg Blue-M tube furnace. An ultra-high purity (UHP) He is used as a carrier gas at a flow of 100 sccm and pressure of 0.1 bar. The flow rate and pressure inside the growth tube are controlled with two needle valves positioned near the two ends of the tube. A stoichiometric mix (1:1 molar ratio) of the precursor materials, cesium bromide (CsBr) and lead(II) bromide (PbBr₂), both of 99.999% purity (Sigma-Aldrich), was used as a source. An alumina boat with the mix of precursors is placed inside the quartz tube at the center of the furnace, and freshly exfoliated muscovite mica used as a substrate is placed 9 cm downstream. In our setup, the position of the tube furnace that defines the temperature profile and the position of the source inside the quartz tube can be both changed with respect to the quartz tube holding a stationary substrate. Importantly, these shifts can be performed finely and *in-situ*, during the growth.

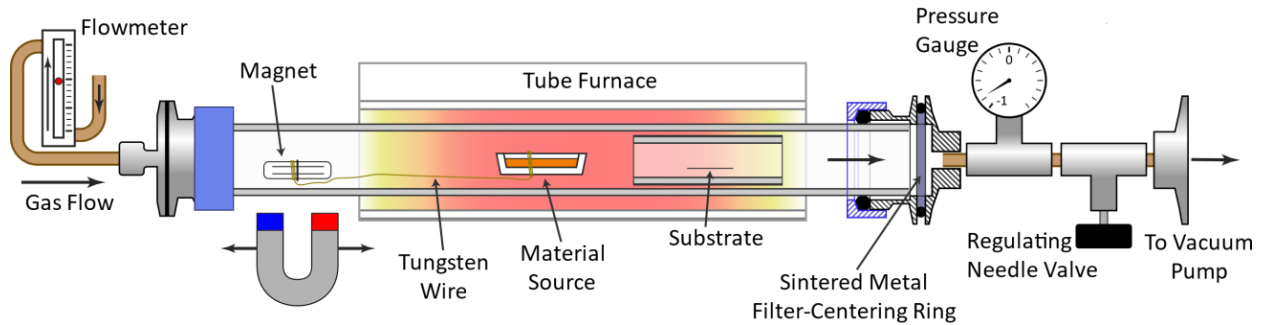

**Figure S1.** A sketch of the home-built apparatus used for the vapor-phase epitaxial growth of CsPbBr₃ films on single-crystal substrates.

The source material is first melted by setting the furnace's temperature to 560 °C, which initiates the heating phase taking 1 h. During this phase, the substrate is also heated to a temperature above the sublimation point of the perovskite and its precursors, thus resulting in the removal of any undesired disordered material that could have been deposited on the substrate during the initial phase of ramping up the temperature. This step is crucial, as it ensures an automatic *in-situ* cleaning of the substrate. After such a cleaning/annealing and stabilizing the temperature, an epitaxial growth of a crystalline CsPbBr₃ on the mica substrate is initiated by slightly reducing the temperature of the substrate achieved by slowly shifting the furnace and the source with respect to the quartz tube by 1 inch in the direction opposite to the gas flow, without changing the set $T$ of the furnace.  A typical growth phase lasts 15 min.  The growth is then abruptly terminated by withdrawing the source from the hot zone of the furnace without changing the position of the substrate with respect to the temperature distribution.  This is achieved with the help of magnets (**Fig. S1**).  After the removal of the material source, the grown CsPbBr$_3$ film starts to gradually sublime.  Shortly after this, the furnace is turned off, and the substrate gradually cools down to room temperature.  It is important that the growth process in our apparatus can be abruptly terminated by withdrawing the source out of the hot zone of the furnace, without changing the position of the substrate relatively to the stabilized temperature profile.  This is crucial for achieving the high-quality films, as well as to control their thickness.  The thickness of the resultant CsPbBr$_3$ film depends on the duration of the growth phase and on the time passed between the extraction of the source and turning off the furnace.

### 2. Helium ion microscopy (HeIM).

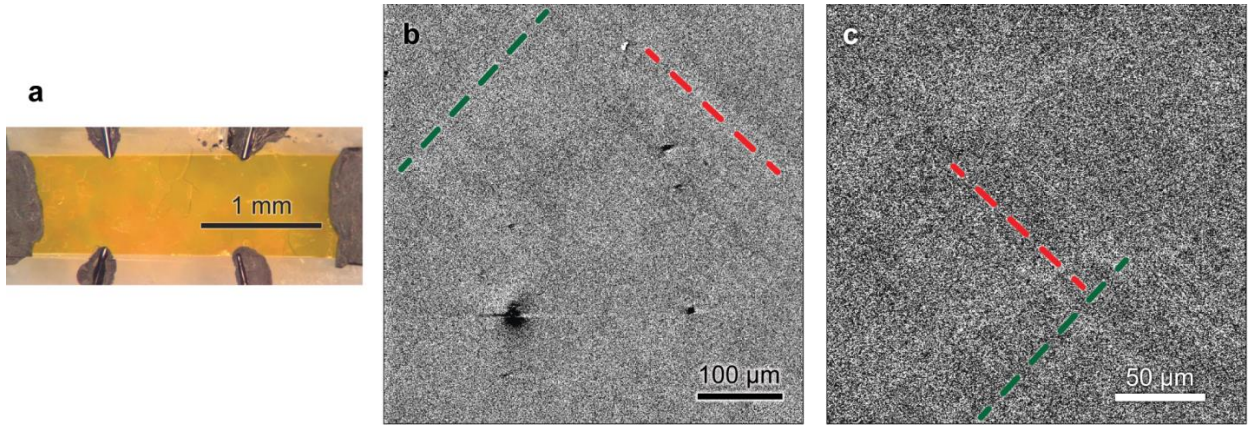

**Figure S2.  Revealing the cross-hatch texture in HeIM images of epitaxial CsPbBr$_3$ films.  (a)** Photograph of a CsPbBr$_3$ device (before deposition of the gate dielectric and the gate), showing a patterned perovskite film (yellow) and graphite contacts (black).  **(b, c)** Magnified HeIM images of the central area of the crystalline channel (presented in **Fig. 1e** of the main text).  Because the cross-hatch texture observed on some of the epitaxial CsPbBr$_3$ films grown on mica is very faint, it was necessary to increase the apparent contrast of these images, which has in turn led to a greater image noise.  The green and red dashed lines are guides to eye showing the orientation of the cross-hatch pattern.

HeIM was used to explore the surface morphology of the epitaxial $CsPbBr_3$ films on various length scales. Of particular use were the observations of the channel of our devices on macroscopic length scales (a typical device with a ~ 2 mm-long crystalline channel is shown in **Fig. S2 a**). HeIM could be used to visualize certain long-range surface features of the crystalline perovskite films. HeIM images of a uniform crystalline surface of our epitaxial $CsPbBr_3$ films usually reveal a very faint *cross-hatch texture* that is most visible at intermediate magnifications. **Figures S2 b** and **c** show the typical surface texture of a single crystalline region of one of the epitaxial $CsPbBr_3$ films recorded in two magnifications. The contrast of these images is enhanced compared to the HeIM image in **Fig. 1e** of the main text. A diagonally aligned texture is visible on these images (highlighted with the green and red dashed lines).

We used a two-dimensional fast Fourier transform (2D-FFT) of large-scale HeIM images to study the long-range crystalline morphology of our films. **Figure 1f** of the main text reveals the preferred orientation of the texture pattern with a very narrow angular distribution. The corresponding scan was taken at a highly ordered single-crystal grain of an epitaxial $CsPbBr_3$ film. In order to compare this result with 2D-FFT analysis of less ordered regions of the films, we have performed a similar analysis of the region near the grain boundary, between two misaligned single crystalline $CsPbBr_3$ grains. **Figure S3a** shows a HeIM scan of such a region containing cracks and other irregularities that might be associated with inhomogeneous strain in this region. A lower crystalline quality and the presence of cracks are typical for regions near grain boundaries. Different single-crystal grains have different mutual orientation of the cracks. Thinner $CsPbBr_3$ films have fewer cracks, suggesting that cracks likely develop during the growth because of the different coefficients of thermal expansion of mica and $CsPbBr_3$.

**Figure S3b** shows the 2D-FFT spectrum of this region, highlighting numerous preferred orientations of the cracks and other features of the surface texture seen on the real-space image in **Fig. S3a**. The green and red dashed lines are added to show the dominant orientations of the cracks in the real-space image and in the corresponding 2D-FFT spectrum.

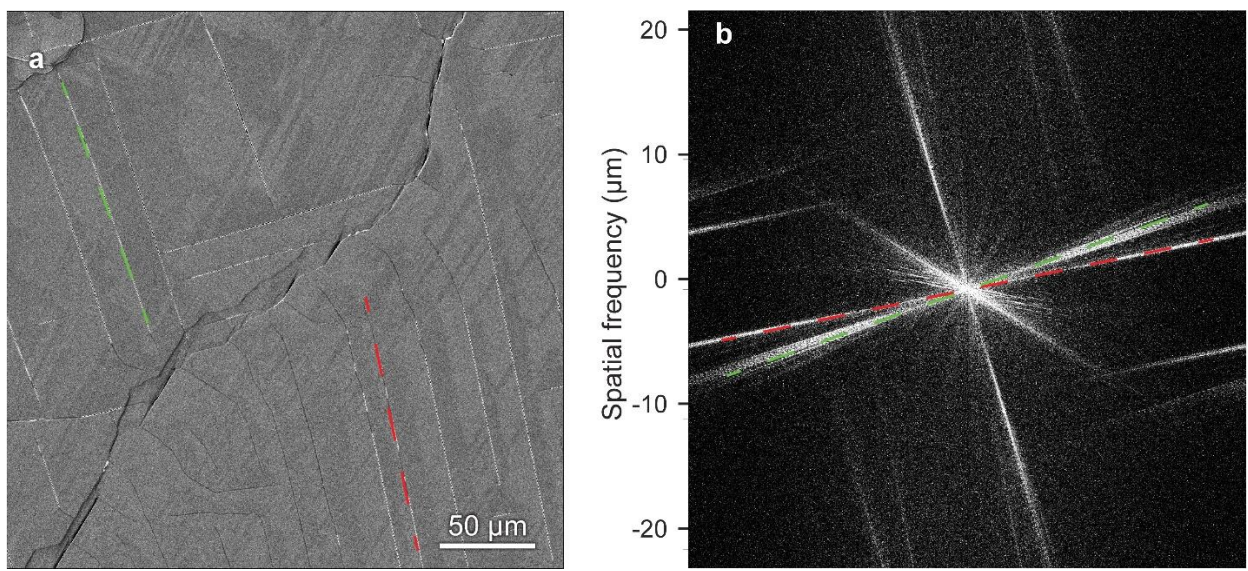

**Figure S3. HeIM imaging and 2D-FFT image processing of a more disordered region of CsPbBr₃ film.** (a) A HeIM image of a CsPbBr₃ film containing a grain boundary (the dark S-shaped crack splitting the image from the top-right to the bottom-left corners) and thin cracks occurring in the two neighboring crystalline grains at a low angle to each other. The green and red dashed lines are added to highlight the orientation of the thin cracks. (b) A 2D-FFT spectrum of the HeIM image shown in the panel (a). The radial bright lines correspond to the preferential orientation of the cracks and other features of the surface texture. The red and green dashed lines, corresponding to the respective lines on the real-space image, show the preferential orientations of the thin cracks.

### 3. X-ray photoelectron spectroscopy (XPS).

The stoichiometry of lead-halide perovskites can vary significantly depending on the growth conditions.[1] To investigate the stoichiometry of our films we have characterized them with XPS. **Table S1** below lists the measured composition of different samples, with all the data normalized to Cs content. The atomic ratios were estimated using the ALTHERMO1 cross-section library. The apparent stoichiometry of all the samples is significantly different from the expected nominal one (that is, Pb:Cs = 1:1, and Br:Cs = 3:1). However, the difference between the as-grown epitaxial films, the *ex-situ* or *in-situ* scratched films, and the *ex-situ* or *in-situ* cleaved stoichiometric crystallized melt is comparable to that observed between different spots probed at the surface of each type of sample. Similar apparent off-stoichiometry has been observed in XPS studies of CsPbBr₃ before.[1-4] Such apparent deviations in the stoichiometry, commonly observed in XPS measurements of perovskites, can be attributed to inaccurate cross-sections used in the calculations of the atomic composition based on the XPS data. It is more important that within the experimental error the surface stoichiometry of our epitaxial CsPbBr$_3$ films (used in the FET fabrication in this work) is the same as that of the bulk CsPbBr$_3$ crystals obtained from a crystallized stoichiometric melt.

**Table S1. The elemental composition of various epitaxial crystalline CsPbBr$_3$ films and bulky (polycrystalline) crystallized melt, as determined from XPS measurements.**

| Sample | Pb/Cs | Br/Cs |
|---|---|---|
| **Epitaxial crystal top surface** | | |
| 1 | 0.54 | 2.17 |
| 2 | 0.49 | 2.13 |
| **Scratched in UHV epitaxial crystal** | | |
| 1 | 0.61 | 2.49 |
| 2 | 0.56 | 2.23 |
| **Stoichiometric polycrystalline melt cleaved in air** | | |
| 1 | 0.55 | 2.14 |
| 2 | 0.59 | 2.32 |
| 3 | 0.61 | 2.17 |
| **Stoichiometric polycrystalline melt cleaved in UHV** | | |
| 1 | 0.60 | 2.38 |
| 2 | 0.55 | 2.23 |
| 3 | 0.56 | 2.33 |

Because in this work we handle all our samples in a regular laboratory air, the surface of our epitaxial CsPbBr$_3$ films shows some contamination with carbon and oxygen. **Figure S4** shows an XPS depth profile calculated from the XPS spectra sequentially taken after a repeated sputtering of a very thin surface layer of one of such samples. Only the first point (at zero depth, $x = 0$), representing the as-grown surface of this CsPbBr$_3$ film, shows detectable amounts of the contaminants. After the very first sputtering step, these contaminants vanish below the detection threshold. Thus, the XPS depth profiling confirms that C and O are only detectable within the surface of the epitaxial films.

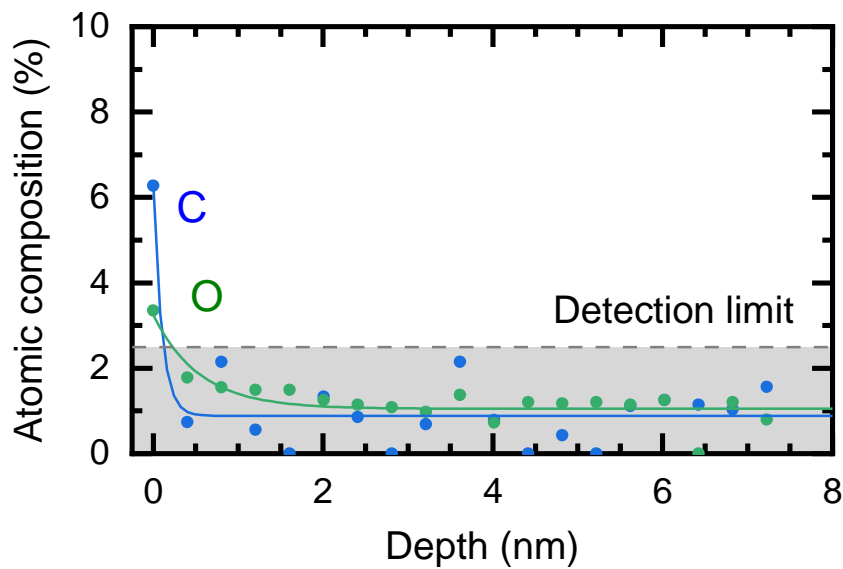

**Figure S4. XPS depth profiling of O and C in an as-grown epitaxial crystalline CsPbBr₃ film handled in air.** The detection limit of XPS (estimated to 2.5 % of the atomic composition) is marked by the horizontal dashed line, below which the atomic composition cannot be reliably determined (grey shade). O and C atoms can be detected with confidence only at the very surface of the as-grown film.

### 4. The effect of poling.

Lead-halide perovskites are known to exhibit a significant *poling effect* - a gradual change of the sample's conductivity occurring in an external electric field. This effect is believed to be of an electro-chemical nature, related to an ionic drift in the material.[5-6]

In our freshly prepared CsPbBr₃ FETs, a gradual modification of the devices' electrical properties also occurs, when a source-drain voltage, $V_{SD}$, is initially applied at room temperature. A noticeable carrier concentration existing in the channel due to the interfacial doping has allowed us to mildly pole the devices at $V_G = 0$ and $V_{SD} = -10$ V (which would normally correspond to a saturation regime in undoped regular semiconductor FETs), yet still avoid driving the device into the saturation regime. The latter helps avoiding the degradation occurring in perovskite FETs in the saturation regime due to a very strong local longitudinal electric field generated in the depletion region. Avoiding such a regime eventually leads to the FET output and transfer characteristics with a very small hysteresis (**Fig. 3** of the main text). Under such a mild poling (performed at $V_G = 0$ and $V_{SD} = -10$ V for 24 h in high vacuum at room temperature), the channel's four-probe conductivity, $\sigma_{4p}$, gradually improves, although the magnitude of the effect depends on the ambient atmosphere and pressure. The FET characteristics of our devices improve after such a pre-poling of the devices. Investigating the details of the poling effect in epitaxial CsPbBr$_3$ FETs, including its dependence on the type of the interface (a particular combination of perovskite and gate dielectric) and atmosphere, is very important and will encompass a series of systematic experiments. Such study is, however, beyond the scope of this work.

### 5. Parylene doping of CsPbBr$_3$.

A positive onset voltage observed in the transfer characteristics of our FETs (**Fig. 3 c** and **d** of the main text) originates from a doping of the channel by the polymeric gate dielectric (*parylene*) grown directly at the surface of CsPbBr$_3$. The corresponding (dark) conductivity is observed in these devices even before the deposition of the gate electrode. A stronger doping effect is observed with parylene-*F*, compared to the milder effect of parylene-*N*, and it is thus likely associated with electron withdrawing properties of parylenes interfaced with CsPbBr$_3$. Similar interfacial doping has been observed in organic semiconductors (e.g., rubrene or tetracene single crystals) coated with a perfluoropolyether (PFPE).[7] Similarly to PFPF, parylene-*F* contains C-F functional groups that might carry a dipole moment. These functional groups can induce an interfacial charge-transfer doping when the polymer is deposited at the semiconductor's surface. In addition, if ordering of these functional groups occurs at the semiconductor's surface, they collectively may result in an electrostatic carrier injection from contacts. Although parylene-*N* does not induce an interfacial doping in rubrene crystals, parylene-*F* does. In CsPbBr$_3$, these effects from the parylenes seem stronger, with both the parylene-*N* and *F* inducing a noticeable surface doping.

### 6. On the background subtraction in the Hall effect measurements of CsPbBr$_3$ FETs.

The base-line conductivity of CsPbBr$_3$ may exhibit a very slow drift and long-term fluctuations occurring in response to abrupt changes in the device's working conditions, such as changes in operating temperature or applied voltages. These drifts and fluctuations can cause the corresponding variations in the parasitic zero-magnetic-field background voltage between the Hall voltage probes (the Hall voltage offset, $V_{\text{Hall-0}}$). To correctly determine the relatively small, true Hall voltage, $V_{\text{Hall}}(B)$, on top of this drifting background, the total measured Hall voltage signal, $V_{\text{Hall}}^{\text{total}}$, was fitted with the following function of two independent variables, $t$ and $B$:

$$V_{\text{Hall}}^{\text{total}} = V_{\text{Hall-0}}(t) + K \cdot B = P_0 + P_1 t + P_2 t^2 + P_3 t^3 + K \cdot B,$$

where $t$ is time, and $P_0$, $P_1$, $P_2$, $P_3$ and $K$ are the fitting parameters. The slowly changing background $V_{\text{Hall-0}}$ is represented by the polynomial. The parameter $K \equiv (V_{\text{Hall}}^{\text{total}} - V_{\text{Hall-0}})/B$ was used in the calculation of the Hall mobility:

$$\mu_{\text{Hall}} = \frac{K}{V_{4\text{p}}} \cdot \frac{D}{W},$$

where $V_{4\text{p}}$ is the longitudinal voltage drop measured by the four-probe voltage contacts, $D$ is the distance between these voltage contacts, and $W$ is the channel width. **Figure S5a** shows an example of the raw $V_{\text{Hall}}^{\text{total}}$ data before subtracting the background and the corresponding fit. **Figure S5b** shows the corresponding data with $V_{\text{Hall-0}}$ subtracted.

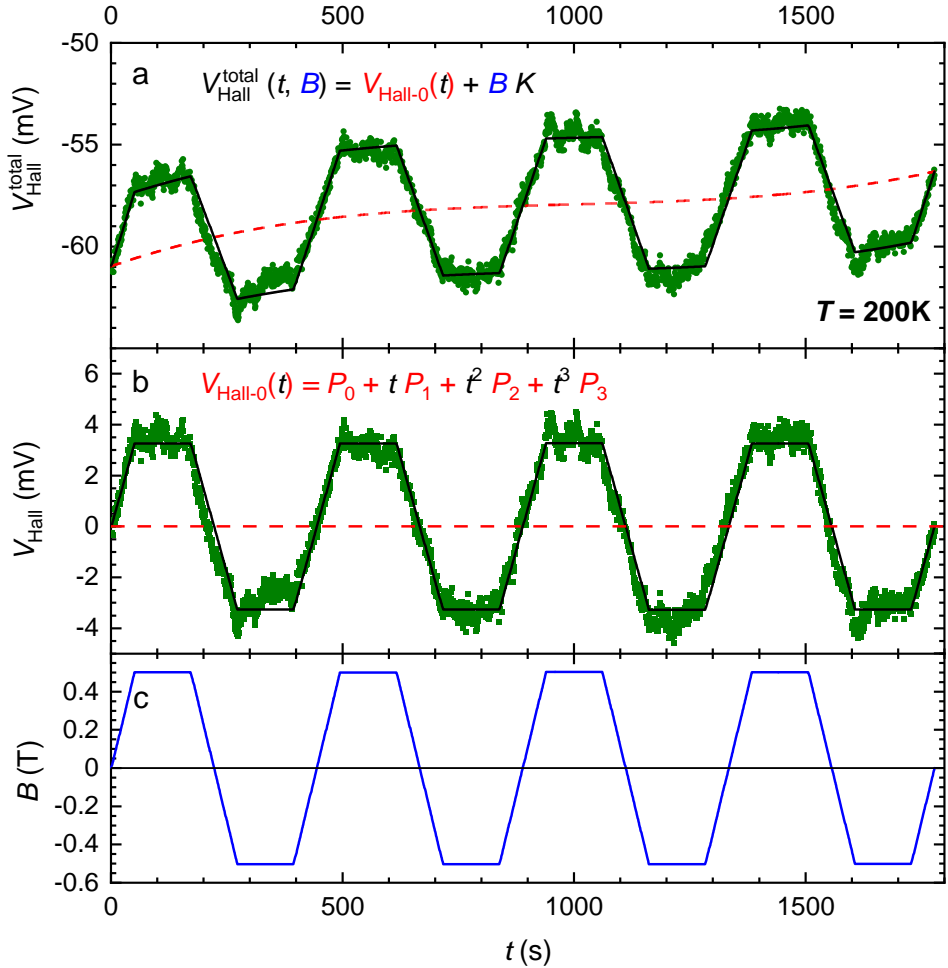

**Figure S5. Subtracting a slowly drifting background (the Hall voltage offset) in Hall measurements.** (a) The raw Hall voltage, $V_{Hall}^{total}$, monitored as a function of time (solid green dots) and a fit as described in this SI (solid black line). The pure drifting background $V_{Hall-0}$, approximated with a polynomial function with 4 parameters $P_n$ ($n$ = 0, 1, 2 and 3) as described in this SI, is also plotted (dashed red line). (b) The Hall voltage data after subtracting the background. (c) The *dc* magnetic field $B$ as slowly swept by the electromagnet. Note, this is a *dc* Hall effect measurement (unlike the *ac* photo-Hall measurement described in the next section): the corresponding Hall voltage is measured by an electrometer, rather than a lock-in amplifier (that is, without a reference to any frequency). The flipping magnetic field is needed as a check for the correct polarity change in the measured Hall voltage.

## 7. The (mis)match between the Hall and FET charge carrier mobilities.

The experimental data on the linear-regime FET and Hall mobilities, $\mu_{FET}$ and $\mu_{Hall}$, in epitaxial CsPbBr3 FETs presented in the main text (**Figs. 3e** and **4**) show that these mobilities could exhibit a noticeable difference, even within the same device. For instance, for the device in **Fig. 3e**, the ratio $\mu_{Hall}/\mu_{FET}$ is about 1.5, while for the device in **Fig. 4**, this ratio is about 1 at room temperature and becomes < 1 at low temperatures. The fact that the difference between $\mu_{Hall}$ and $\mu_{FET}$ slightly varies from device to device suggests that it is device specific and is not of the intrinsic nature. It likely originates from sample-specific morphological defects at the surface of CsPbBr$_3$ channel (e.g., the presence and specific distribution of nano and micro cracks in the channel) that are hard to control/identify visually.

However, besides these extrinsic effects, there could be two fundamental mechanisms contributing to the difference between $\mu_{Hall}$ and $\mu_{FET}$ observed in some of our CsPbBr$_3$ FETs. One of these could be associated with the presence of a relatively small density of hoping carriers, originating from a small residual density of shallow "tail" states (shallow traps), that coexist with band carriers in the FET's accumulation channel, as described by a model recently developed by Yi *at al*.[8] Indeed, the behavior shown in **Fig. 3e** of the main text ($\mu_{Hall} > \mu_{FET}$, with both mobilities being carrier-density independent) is reminiscent of the transport regime denoted in the model by Yi *et al.* as the "weak hopping contribution" (case (**c**), Equation 15 of Ref. [8]). In this regime, the dominant fraction of carriers are band carriers characterized with a high drift mobility, with the rest of the carriers moving in localized tail states (below the mobility edge) via hopping with a very low carrier mobility. This regime describes the materials that are *nearly* 100% intrinsic (trap-free) band semiconductors. The analysis of the model shows that in this regime, $\mu_{Hall}$ is very close to the intrinsic band-carrier mobility of the ideal crystal, and the ratio $\mu_{FET}/\mu_{Hall}$ is equal to the fraction of the band carriers among the total carrier population in the channel (~ 2/3 in the case presented by **Fig. 3e** of the main text), while both $\mu_{Hall}$ and $\mu_{FET}$ would exhibit a similar band-like temperature dependence (with d$\mu$/d$T$ < 0). The case of closely matching $\mu_{Hall}$ and $\mu_{FET}$ (as in **Fig. 4** of the main text at high $T$) corresponds in the model by Yi *et al.* to a negligible density of hopping carriers. Overall, this behavior is consistent with a high-quality, (nearly) disorder-free, crystalline CsPbBr$_3$ films with (nearly) trap-free perovskite/dielectric interface in our perovskite FETs.

Another mechanism that can potentially contribute to $\mu_{Hall}/\mu_{FET} \neq 1$ is based on a carrier scattering off charged impurities: scattering can give rise to some difference between $\mu_{Hall}$ and $\mu_{FET}$.[9] Such defects can be generated in perovskites at varied quantities by poling. All these mechanisms however are not well understood in perovskites, and the high-performance perovskite FETs developed in this work could serve as an experimental tool for their systematic investigation.

### 8. Formation of cracks in CsPbBr3 films at low temperatures.

During the measurements at cryogenic temperatures, all our devices based on epitaxial single-crystal CsPbBr$_3$ films grown on mica (including the FETs and (ungated) photoconductors) exhibited cracking. **Figure S6** shows a HeIM image of a single crystalline region of the film containing a typical crack developed during the temperature cycling. Such cracks are usually narrower than those formed during the film growth (see, for example, **Fig. S3a** above, and **Fig. 1b** of the main text).

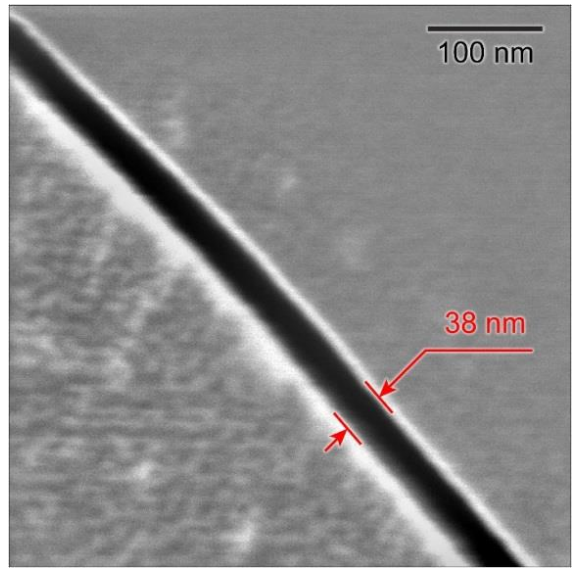

**Figure S6.** A HeIM image of a crack formed in an epitaxial CsPbBr$_3$ film on mica as the result of a temperature cycling of the sample between room temperature and 20 K (that is, after first cooling it to 20 K and then warming it up to room temperature). Typical widths of such cracks are in the range 15 –

50 nm. The formation of these cracks can be detected *in-situ* by monitoring the sample's conductivity (as shown in **Fig. S7** below): sudden drops in the conductivity that start occurring below ~ 200 K (with most of the occurrences below 100 K) indicate formation of cracks.

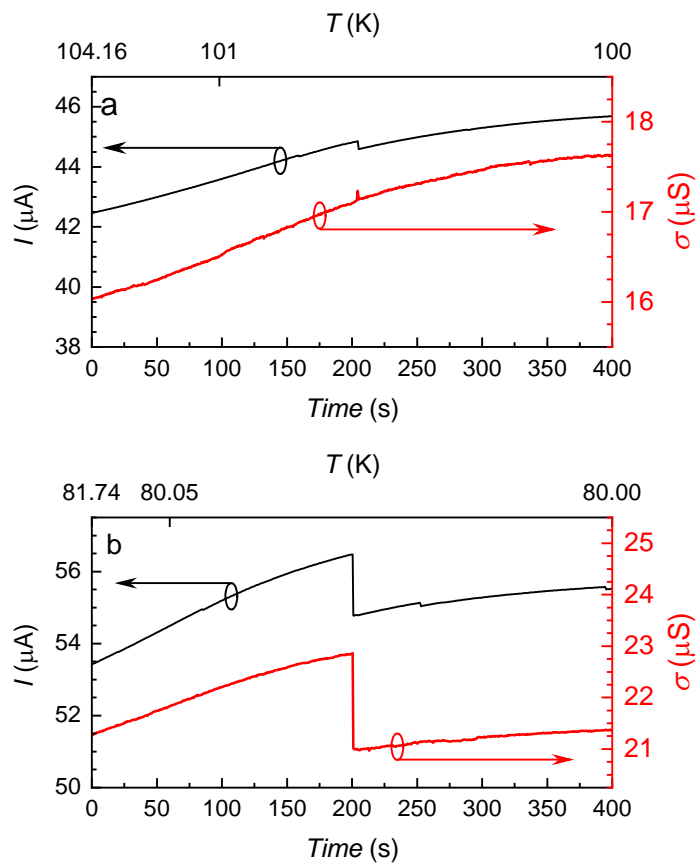

**Figure S7. Monitoring the effect of low-temperature crack formation on the conductivity of epitaxial CsPbBr$_3$ films on mica.** (a) The longitudinal current $I$ (solid black line) and the corresponding four-probe conductivity $\sigma$ (solid red line) measured at around 100 K, when a relatively small crack is formed outside of the central region of the channel probed by the voltage contacts. The crack has formed right below 101 K, which caused the step-wise drop in $I$. No effect of the crack on $\sigma$ is observed in this case. (b) Similar measurements at around 80 K, when another crack has apparently formed within the central region of the channel, causing significant drops in both the current and the four-probe conductivity.

We have first "observed" these cracks in real time by monitoring the sample's conductivity on cooling: formation of the cracks results in small step-wise discontinuities (sudden drops) in the measured current.  Depending on the exact position of the crack in the channel and their size/length they may or may not affect the four-probe conductivity of the film used to evaluate the FET mobility.  Very small cracks start forming at temperatures just below 200 K.  In the range of 100 - 200 K, those small cracks usually do not noticeably affect the longitudinal current and the four-probe voltage and are thus benign (that is, not having an influence on the extracted charge transport parameters).  Below 100 K, however, the effect of cracks on the longitudinal conductivity becomes significant.  **Figure S7** shows typical stepwise drops in the measured current and the corresponding conductivity, as the CsPBr$_3$ film on mica is cooled through 100 and 80 K.  The drop occurred around 100 K (**Fig. S7a)** does not affect the four-probe conductivity, because, evidently, the corresponding crack has formed outside of the central region of the channel probed by the voltage contacts.  Unlike the crack at 100 K, the crack that has occurred around 80 K did lead to a drop in the four-probe conductivity of the sample (**Fig. S7b**).  Below 50 K, all the studied devices were developing more serious cracks that were abruptly reducing the channel's current 2-3-fold, making further measurements unreliable.

### 9.  Measurements of a photo-Hall effect in epitaxial crystalline CsPbBr3 films.

The charge transport in FET geometry is restricted to a very thin layer of the semiconductor adjacent to the interface with the gate dielectric.  Thus, properties of the semiconductor-dielectric interface could significantly affect the charge transport efficiency in FETs.  The morphology and purity of the top surface of epitaxial CsPbBr$_3$ films, as well as the carrier scattering off interfacial phonons, might become the factors limiting the charge carrier mobility, $\mu_{FET}$, of FETs.  Hence, experimental $\mu_{FET}$ does not necessarily reflect the intrinsic charge transport properties of the semiconductor's bulk.  To compare the interfacial carrier mobility as measured in our FETs with the characteristic bulk mobility of single-crystal CsPbBr$_3$, we have carried out photo-Hall effect measurements in ungated devices based on the same epitaxial CsPbBr$_3$ films.

**Figure S8a** shows a sketch of the photo-Hall effect measurement configuration using an ungated epitaxial CsPbBr$_3$ device under a *cw* illumination.  In this case, we used a highly sensitive *ac*-Hall measurement technique, originally developed by our group for organic semiconductor FETs,[10] and lately used for the analogous photo-Hall measurements of solution-grown bulk perovskite crystals.[11] Here, we used this methodology to reliably measure the photo-Hall voltage, $V_{\text{photo-Hall}}$, in our epitaxial perovskite films, which can be used to estimate the bulk carrier mobility of these samples. **Figure S8b** is a microphotograph of one of our photo-Hall devices used for the photo-Hall measurements. In essence, the difference between the FET and the photo-Hall device structures is that the latter lacks the gate dielectric and the gate electrode. In this configuration, the film's conductivity is governed by the photo-generated carriers. **Figure S8c** shows the r.m.s. of a photo-Hall voltage signal, $V_{\text{photo-Hall}}$, detected with a lock-in amplifier tuned into the frequency of the *ac-B* field (r.m.s. *B* = 0.23 T, *f* = 0.2 Hz), together with the corresponding longitudinal photocurrent, *I*, whose polarity is intentionally switched back and forth. Such a measurement mode helps to correctly account for the parasitic Faraday induction electromotive force (Faraday's EMF) that can be generated by the changing magnetic field in the circuit loop formed by the wires connected to the Hall voltage probes.

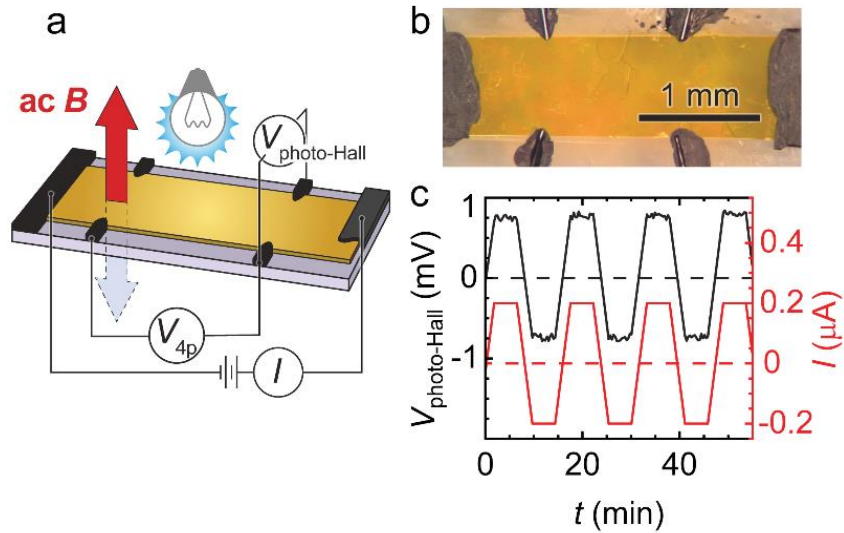

**Figure S8. The photo-Hall effect measurements in (ungated) epitaxial single-crystal CsPbBr₃ films.** (a) A sketch of the measurements performed under a uniform *cw* illumination via a high-resolution *ac*-Hall technique that uses a uniform *ac* magnetic field *B*, a longitudinal excitation *dc* current *I*, and a lock-in detection of the corresponding *ac* photo-Hall voltage, $V_{\text{photo-Hall}}$. (b) An optical microphotograph of a CsPbBr$_3$ single-crystal device with graphite contacts in a four-probe/Hall-bar geometry. (c) The r.m.s. of $V_{\text{photo-Hall}}$ signal (black solid line) measured as the longitudinal photocurrent (red solid line) flows along the film under *cw* illumination. The polarity of the current is intentionally switched back and forth, which allows to correctly account for the parasitic Faraday induction EMF.